\documentclass[a4paper]{article}

\usepackage{INTERSPEECH2022}
\usepackage{multirow}
\usepackage{subfig}
\usepackage{booktabs}
\usepackage{url}
\title{Improving Mandarin Prosodic Structure Prediction with Multi-level Contextual Information}
\name{
Jie Chen$^{1,\dagger,\ddagger}$, Changhe Song$^{1,\dagger}$, Deyi Tuo$^{3}$, Xixin Wu$^{4}$, Shiyin Kang$^{2,*}$, \\ Zhiyong Wu$^{1,4,*}$, Helen Meng$^{1,4}$
\thanks{$\ddagger$ Work performed when Jie Chen was interning at Huya Inc.}
\thanks{$\dagger$ Equal contribution. * Corresponding author. } 
}
\address{
    $^1$ Shenzhen International Graduate School, Tsinghua University, Shenzhen, China\\
    $^2$ XVerse Inc., Shenzhen, China\quad$^3$ Huya Inc., Guangzhou, China\\
    $^4$ The Chinese University of Hong Kong, Hong Kong SAR, China
 }
\email{
\{chenjie20, sch19\}$@$mails.tsinghua.edu.cn,
\{wuxx, hmmeng\}$@$se.cuhk.edu.hk,
tuodeyi$@$huya.com,
zywu$@$sz.tsinghua.edu.cn,
kangshiyin$@$xverse.cn
}

\begin{document}

\maketitle

\begin{abstract}
For text-to-speech (TTS) synthesis, prosodic structure prediction (PSP) plays an important role in producing natural and intelligible speech.
Although inter-utterance linguistic information can influence the speech interpretation of the target utterance, previous works on PSP mainly focus on utilizing intra-utterance linguistic information of the current utterance only.
This work proposes to use inter-utterance linguistic information to improve the performance of PSP.
Multi-level contextual information, which includes both inter-utterance and intra-utterance linguistic information, is extracted by a hierarchical encoder from character level, utterance level and discourse level of the input text.
Then a multi-task learning (MTL) decoder predicts prosodic boundaries from multi-level contextual information.
Objective evaluation results on two datasets show that our method achieves better F1 scores in predicting prosodic word (PW), prosodic phrase (PPH) and intonational phrase (IPH). 
It demonstrates the effectiveness of using multi-level contextual information for PSP. 
Subjective preference tests also indicate the naturalness of synthesized speeches are improved\footnote{Synthesized speech samples at: https://thuhcsi.github.io/mlc-PSP/}.
\end{abstract}
\begin{figure*}[!htb]
\centering
\subfloat[Proposed model]{
\includegraphics[width=0.3\linewidth]{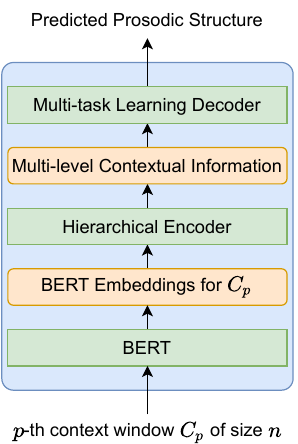}
\label{fig:overview}
}
\quad
\subfloat[Hierarchical encoder]{
\includegraphics[width=0.3\linewidth]{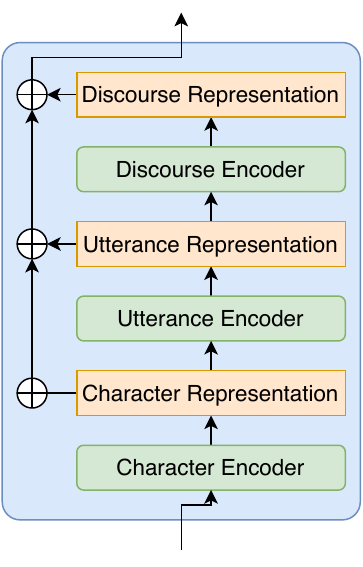}
\label{fig:mlencoder}
}
\quad
\subfloat[Utterance/discourse encoder]{
\includegraphics[width=0.3\linewidth]{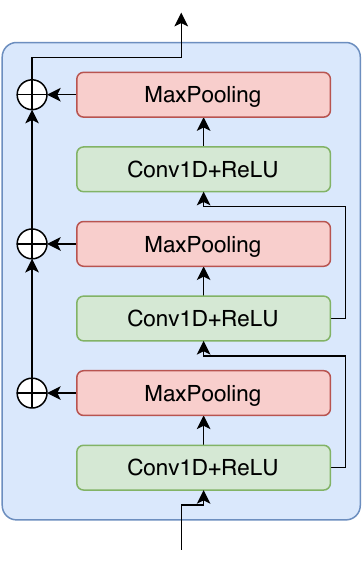}
\label{fig:sde}
}
\caption{The overall architecture for proposed model.}
\label{fig:model_structure}
\end{figure*}

\section{Introduction}

In human conversation, speakers tend to insert different levels of breaks at appropriate positions according to the semantic meaning of the text and the intentions to be conveyed. 
Such phenomenon is called prosodic phrase grouping \cite{Tseng02,peng04}.
The inserted breaks chunk the utterance text into different prosody constituents including PW, PPH and IPH, which can be organized in a hierarchical way as the prosodic structure \cite{chu2001locating}.
In TTS synthesis, PSP aims to predict the aforementioned hierarchical prosodic structure from the input text, which is cruical to the naturalness and intelligibility of the synthesized speech \cite{xutan2021}. 

The PSP task is generally regarded as a sequence-to-sequence based classification problem to predict whether there is a prosody break (i.e. PW, PPH or IPH boundary) after each character of the input text. 
Previous works have carried out relevant studies on not only feature engineering but also model structures for the PSP task. 
For feature engineering, early studies have investigated several human designed features and trainable word embedding \cite{zheng2016improving}, which are later replaced by bidirectional encoder representation from Transformers (BERT) \cite{du2019prosodic, zhang2020unified} to provide richer linguistic information.
Syntactic features such as phrase structure tree and dependency tree also benefit PSP \cite{Chen2010ImprovingPP,Che2014ImprovingMP,Zhang2016MandarinPP}.
With the development of neural network based
methods,
%PSP model, 
an end-to-end PSP model BLSTM-CRF \cite{zheng2018blstm, pan2019mandarin} was proposed which integrates bidirectional long short-term memory (BLSTM) recurrent neural network (RNN) \cite{ding2015automatic,huangyuchen17} and conditional random field (CRF) \cite{levow2008automatic,qian2010automatic}.
Maximum entropy model is also employed in PSP \cite{li2004chinese}.
RNN is further replaced with multi-head self-attention to better model long-term dependencies \cite{du2019prosodic,lu2019self}. 
Moreover, separate classifiers for PW, PPH and IPH are integrated under a framework \cite{pan2019mandarin}, which utilizes dependencies between three tasks and improves overall prediction performance.

%zywu-0327
%这部分重新组织
% [re zywu-0327 from sch] 来源于Paragraph Prosodic Patterns to Enhance Text-to-Speech Naturalness的intro第一段
However, current state-of-the-art PSP models still lack of ability to fully model naturalness and expressiveness to resemble human speech, especially in multi-sentential discourse \cite{peiro2018paragraph}. 
% 给出文章改进范围：multi-sentential discourse
Several works in literature have proved the existence of supra-sentential prosody patterns in discourse segments, e.g. the increase in numbers of breaks, pause and lengthening before sentential boundary due to the availability of neighboring semantic information \cite{kreiman1982perception, gussenhoven1992perceptual}, the declination of pitch through both intra- and inter-sentential units \cite{de2015pitch}, etc. 
Thus, introducing inter-utterance linguistic information from adjacent utterances can have the potential to benefit PSP. 
% 给出具体例证
%To consider contextual information from neighboring sentences or utterances, for the task of acoustic modeling in TTS, there have been some studies that adopt hand designed features or text embeddings of adjacent utterances to provide inter-utterance linguistic information for predicting acoustic features directly \cite{KlimkovNMPBMD17,le2016marytts,sawada2016nitech,liao2018ntut,xu2020improving}. 
Some studies in TTS adopt hand designed features or text embeddings of adjacent utterances to provide inter-utterance linguistic information for predicting acoustic features directly \cite{KlimkovNMPBMD17,le2016marytts,sawada2016nitech,liao2018ntut,xu2020improving}. 
%In \cite{KenterWCCV19}, a hierarchically structured conditional variational auto-encoder has been proposed to capture the linguistic hierarchy incorporated in the input utterance to improve the prosody of generated speeches.
Previous work also suggests that effective incorporation of inter-utterance linguistic information can improve automatic speech recognition \cite{sun2021transformer}.

In this work, we propose to use multi-level contextual information, which includes
not only intra-utterance linguistic information of the current utterance
but also inter-utterance linguistic information from adjacent utterances,
to improve the performance of PSP.
%To integrate the multi-level contextual information into PSP, we propose a novel PSP model based on encoder-decoder architecture \cite{sutskever2014sequenceto} which has a hierarchical encoder and a MTL decoder.
To integrate the multi-level contextual information into PSP, we propose a novel PSP model based on encoder-decoder architecture which has a hierarchical encoder and a MTL decoder.
The hierarchical encoder consists of a character encoder, an utterance encoder and a discourse encoder and is responsible for extracting multi-level contextual information from the input text.
The character encoder is composed of several Transformer encoder blocks \cite{vaswani2017attention}.
The utterance encoder and the discourse encoder employ multi-layer convolutional neural network (CNN) \cite{kim2014convolution}.
A group of adjacent utterances in the same context are sent to the character encoder to produce character representations, which are further processed by the utterance encoder to get utterance representations. 
Then, discourse encoder capturing inter-utterance linguistic information from utterance representations produces a discourse representation. 
Finally, these three-level representations are concatenated together as the  multi-level contextual information which is fed into a MTL decoder to predict PW, PPH and IPH boundaries. 

Results from objective experiments show that our model achieves better prediction performance in PW, PPH and IPH subtasks, which shows the effectiveness of using multi-level contextual information for the PSP task.
Furthermore, ABX preference tests also indicate our method can derive better naturalness of synthesized speech.

\section{Methodology}
\label{s1:method}

%zywu-0327
%这段英文写的不好，需要改进
%图引用时要加上括号，比如Fig.1(a) ok
%Our proposed model is aimed at predicting the prosodic structure conditioned on multi-level contextual information.  
%BERT representations of the utterances are fed into the hierarchical encoder to get multi-level contextual information.
%Then, multi-level contextual information is sent to the MTL decoder to finish the prediction task. 
Fig.\ref{fig:model_structure}\subref{fig:overview} shows the overall architecture of our proposed model including a pretrained BERT, a hierarchical encoder and a MTL decoder.

\subsection{Hierarchical encoder}
\label{ssec1:metho_encoder}
The hierarchical encoder belongs to PSP models with self-attention \cite{du2019prosodic,lu2019self} and consists of a character encoder, an utterance encoder and a discourse encoder.
Fig.\ref{fig:model_structure}\subref{fig:mlencoder} shows its architecture.
Assume a Chinese document $D=\{s_1,s_2,s_3,...,s_N\}$ has $N$ utterances, where $s_i$ is the $i$-th utterance in the original document. 
A sequence of $n$ utterances from $s_p$ to $s_{p+n-1}$ form the $p$-th context window of size $n$ in $D$, denoted by $C_p=\{s_p,s_{p+1},s_{p+2},...,s_{p+n-1}\}$.
Let $C_{pj}$ be the $j$-th utterance in $C_p$, and $C_{pjt}$ be the $t$-th Chinese character of $C_{pj}$. 
For $C_p$, all utterances will be embedded with a character-level BERT to get the feature $\text{B}_p\in \mathbb{R}^{n\times l \times d}$, where $l$ is the length of the padded utterance, $d$ is the dimension of the BERT feature. 
Then $\text{B}_p$ is passed to the hierarchical encoder to extract multi-level contextual information.

\subsubsection{Character encoder}
\label{sssec2:metho_encoder_character}
%zywu-0327
%给出Transformer的引用 ok
The character encoder utilizes Transformer encoder blocks \cite{vaswani2017attention} to extract character representations, 
in which the self-attention mechanism 
%provided by Transformer encoder block 
helps learn long-range dependencies inside each utterance by connecting two arbitrary characters directly regardless of their distance \cite{lu2019self}.
Positional encodings \cite{vaswani2017attention} are added to $\text{B}_{pj}$ to assist Transformer encoder blocks distinguish different positions of the input sequence.
The character representation $\text{CR}_{pj} \in \mathbb{R}^{l \times d}$ of the utterance $C_{pj}$ is produced by the character encoder. 
That is:
\begin{equation}
\text{CR}_{pj}=\text{CharacterEncoder}(\text{B}_{pj}+\text{PE})
\end{equation}
where $\text{PE}$ is the positional encoding.

\subsubsection{Utterance encoder}
\label{sssec3:metho_encoder_utterance}
The utterance encoder is responsible for aggregating the information from one utterance into a single vector, which is shown in Fig.\ref{fig:model_structure}\subref{fig:sde}.
For this encoder, we use multi-layer CNN.
%CNN with an increasing receptive field can better extract local-global features and reveal high-level interactions between characters \cite{feng2017aconvolutional}. 
%For utterance $C_{pj}$, its character representation $\text{CR}_{pj}$ is processed by a stack of convolution layers. 
The character representation $\text{CR}_{pj}$ of utterance $C_{pj}$ is processed by a stack of 1-D convolution layers. 
Assume we have a total of $m$ convolution layers.
The $r$-th convolution layer $\text{CONV}_r$ has $k_r$ convolution kernels. %whose width is the same as the dimension of $\text{CR}_{pjt}$.
All convolution layers have ReLU activation function.
The output $Y_r$ of the $r$-th convolution layer is:
\begin{equation}
Y_r=\text{ReLU}(\text{CONV}_r(X_r))
\end{equation}
where $X_1=\text{CR}_{pj}$, $X_{r+1}=Y_r$, and $Y_r\in \mathbb{R}^{l\times k_r}$ is the output of the $r$-th convolution layer.
The output of each convolution layer is passed to a max-over-time pooling operation \cite{kim2014convolution} to capture the most important feature. Finally, the outputs of the pooling layers are concatenated to produce the utterance representation $\text{UR}_{pj}$ of the utterance $C_{pj}$:
\begin{equation}
\text{UR}_{pj}=[\text{MP}(Y_1),~\text{MP}(Y_2),...,~\text{MP}(Y_m)]
\end{equation}
where $[\cdot]$ is the concatenating operation, $\text{MP}$ is the max-over-time pooling operation, $\text{UR}_{pj} \in \mathbb{R}^{\sum_{i=1}^{m}{k_i}}$.

\subsubsection{Discourse encoder}
\label{sssec4:metho_encoder_context}

%zywu-0327
%首先说明Discouse encoder的目的 ok
%最后解释q ok
The discourse encoder, which shares the same architecture as the utterance encoder in Section \ref{sssec3:metho_encoder_utterance}, is responsible for producing discourse representations capturing inter-utterance linguistic information.
Total $n$ utterance representations $\text{UR}_p \in \mathbb{R}^{n \times \sum_{i=1}^{m}{k_i}}$ of $C_p$ are fed into the discourse encoder
%, which has the same structure as utterance encoder in Section \ref{sssec3:metho_encoder_utterance}. 
%Finally, $\text{SR}_p$ is processed to get 
to derive
a single vector as the discourse representation $\text{DR}_p \in \mathbb{R}^q$,
where $q$ is the sum of numbers of convolution kernels in discourse encoder.

\subsection{Multi-task learning decoder}
\label{ssec2:metho_decoder}

For decoder, MTL framework is adopted, where the prediction of PW, PPH and IPH tags are regarded as three different but related subtasks of PSP. 
Each subtask has separate gated recurrent unit (GRU) network. 
To predict the prosodic boundary tag after character $C_{pjt}$, the aforementioned character, utterance and discourse representations  $\text{CR}_{pjt}$, $\text{UR}_{pj}$ and $\text{DR}_p$ are concatenated as the multi-level contextual information $\text{MLCI}_{pjt}$ of $C_{pjt}$.
%\begin{equation}
%\text{MLCI}_{pjt}=[\text{CR}_{pjt},\text{UR}_{pj},\text{DR}_p]
%\end{equation}

The prediction of PW tag for character $C_{pjt}$ doesn't depend on other subtasks, so $\text{MLCI}_{pjt}$ is directly sent to PW GRU to produce its hidden state 
$\text{H}_{pjt}^{\text{PW}}$:
\begin{equation}
\text{H}_{pjt}^{\text{PW}}=\text{GRU}(\text{MLCI}_{pjt})
\end{equation}

For PPH GRU, in addition to  $\text{MLCI}_{pjt}$, it  also accepts $\text{H}_{pjt}^{\text{PW}}$ as part of its input to produce its hidden state $\text{H}_{pjt}^{\text{PPH}}$: %, that is:
\begin{equation}
\text{H}_{pjt}^{\text{PPH}}=\text{GRU}([\text{MLCI}_{pjt},\text{H}_{pjt}^{\text{PW}}])
\end{equation}
For IPH GRU, it accepts not only $\text{MLCI}_{pjt}$ but also hidden states from PW and IPH GRUs, and generates its hidden state $\text{H}_{pjt}^{\text{IPH}}$ as:
\begin{equation}
\text{H}_{pjt}^{\text{IPH}}=\text{GRU}([\text{MLCI}_{pjt},\text{H}_{pjt}^{\text{PW}},\text{H}_{pjt}^{\text{PPH}}])
\end{equation}

%zywu-0327
To predict whether there is a PW, PPH or IPH boundary break after character $C_{pjt}$, three feedforward neural networks (FNNs) with softmax activation function are adopted for the three subtasks respectively, each of which accepts the hidden state of corresponding GRU as input.
Summation of the losses from three subtasks is used as the total loss for optimization.

By conditioning the IPH tag prediction on PPH and PW subtasks, and conditioning the PPH tag prediction on PW subtask, we provide richer information from other subtasks and model the hierarchical dependencies between PW, PPH and IPH, which can improve the overall performance.

\section{Experiments}
\label{sec:experiment}

\subsection{Datasets}
\label{ssec1:datasets}

As there is no public dataset for PSP, we prepared two datasets to validate the performance of our proposed model in different scenarios.
The first dataset is transcribed from a massive open online courses (MOOC) and  prosodic structures are manually labeled according to the corresponding audios. 
%The style of the MOOC is close to the spoken language. 
The second dataset is manually labeled from the recordings of an audiobook, and its text contains both dialogue and narration.
%The language of audiobook dataset is more formal compared to MOOC dataset.
Both datasets are divided into training and test sets with ratio 9:1.
The statistics of the datasets is shown in Table \ref{table:statistics_datasets}.

%zywu-0327
%感觉没必要说，所以注释掉这一段
%Similar to \cite{pan2019mandarin}, we extract mixed label sequences for PW, PPH and IPH from the labeled samples. Then label sequences of PW, PPH and IPH are obtained from mixed label sequences. During this procedure, PPH and IPH labels are added to PW label sequences, and IPH labels are added to PPH label sequences.

\begin{table}
\centering
\caption{Statistics of the datasets.}
\label{table:statistics_datasets}
\begin{tabular}{@{}ccccccc@{}}
\toprule
 & \textbf{Dataset}            &  & \textbf{Type} &  & \textbf{Count} &  \\ \midrule
 & \multirow{5}{*}{MOOC}       &  & utterance      &  & 4,772          &  \\
 &                             &  & character     &  & 160,719        &  \\
 &                             &  & PW            &  & 35,096         &  \\
 &                             &  & PPH           &  & 13,239         &  \\
 &                             &  & IPH           &  & 12,362         &  \\ \midrule
 & \multirow{5}{*}{audiobook} &  & utterance      &  & 11,078         &  \\
 &                             &  & character     &  & 441,819        &  \\
 &                             &  & PW            &  & 73,333         &  \\
 &                             &  & PPH           &  & 46,445         &  \\
 &                             &  & IPH           &  & 43,777         &  \\ \bottomrule
\end{tabular}
\end{table}

\subsection{System setup}
%\subsection{System built}
% 所有模型都基于MTL框架
% 使用BERT作为输入
% 学习率，batch size，优化算法adam。
%In this paper, all PSP models are optimized in MTL framework. 
%In this paper, the input of all models is from a character-level BERT \footnote{https://huggingface.co/bert-base-chinese}, which is not finetuned in our experiments.

%zywu-0327
%这里的论文写作也需要提升
%既然BERT已经provides linguistic features了，
%那我们的方法所提的inter- and intra-utterance linguistic features是什么呢？
Two representative models in recent years are selected for comparison.
All models use pre-trained Chinese character-level BERT\footnote{https://huggingface.co/bert-base-chinese} embeddings as input.
The parameters of BERT are frozen.

\begin{itemize}

\item \textbf{BLSTM-CRF}: 
A baseline model uses BLSTM as encoder and CRF as decoder and has similar architecture to \cite{zheng2018blstm}.
MTL framework is adopted, where the prediction of IPH tag is conditioned on the predicted binary PPH tags and binary PW tags, and the prediction of PPH tags is conditioned on binary PW tags. The value one of the binary tag indicates there's a prosodic boundary after $C_{pjt}$ and zero otherwise.

\item \textbf{Transformer}: 
%This model is designed for ablation study, which has the same architecture as Proposed but no utterance encoder or discourse encoder. It has similar architecture to \cite{lu2019self}.
A baseline model has similar architecture to \cite{lu2019self}. Different from Proposed, it doesn't have utterance encoder or discourse encoder.

\item \textbf{Proposed}: 
Our proposed model.
%, and the detailed architecture is described in Section \ref{s1:method}.
The character encoder has 2 Transformer encoder blocks, and each block has 4 attention heads. 
The dimension of input to the character encoder is 768 and the dimension of feedforward network is 2048.
The utterance encoder has 3 convolution layers whose kernel sizes are 3, and numbers of kernels for each layer are 128, 64, 64 respectively.
The discourse encoder has the same structure as the utterance encoder, but numbers of kernels for each layer are halved.
The size of the context window (i.e. $n$) is 8.%, and the proposed model will predict prosodic boundaries for this group of 8 utterances at the same time.

\end{itemize}

\begin{table}
\caption{
Performance of the compared models on MOOC and audiobook datasets. 
When the utterance encoder and the discourse encoder of Proposed are removed, Proposed becomes Transformer. 
* means the MTL framework is removed from the corresponding model. }
\label{table:experiment_main_table}
\begin{tabular}{@{}llllc@{}}
\toprule
\multicolumn{1}{c}{\multirow{2}{*}{\textbf{Model}}} &
  \multicolumn{3}{c}{\textbf{F1 score}} &
  \multirow{2}{*}{\textbf{Dataset}} \\ \cmidrule(lr){2-4}
\multicolumn{1}{c}{} &
  \multicolumn{1}{c}{\textbf{PW}} &
  \multicolumn{1}{c}{\textbf{PPH}} &
  \multicolumn{1}{c}{\textbf{IPH}} &
   \\ \midrule
Transformer &
  \multicolumn{1}{c}{93.19\%} &
  \multicolumn{1}{c}{72.97\%} &
  \multicolumn{1}{c}{76.50\%} &
   \\
BLSTM-CRF &
  95.08\% &
  74.24\% &
  77.70\% &
   \\
Proposed & \multicolumn{1}{c}{\textbf{95.35\%}} & \multicolumn{1}{c}{\textbf{75.15\%}} & \multicolumn{1}{c}{\textbf{77.89\%}} & \multirow{2}{*}{MOOC} \\ \cmidrule(r){1-4}
Transformer* &
  93.29\% &
  73.20\% &
  76.73\% &
   \\
BLSTM-CRF* &
  94.97\% &
  74.11\% &
  76.52\% &
  \multicolumn{1}{l}{} \\
Proposed* &
  \textbf{95.09\%} &
  \textbf{75.22\%} &
  \textbf{77.48\%} &
  \multicolumn{1}{l}{} \\ \midrule
Transformer &
  93.93\% &
  86.87\% &
  87.67\% &
   \\
BLSTM-CRF &
  \multicolumn{1}{c}{94.41\%} &
  \multicolumn{1}{c}{86.75\%} &
  \multicolumn{1}{c}{87.71\%} &
   \\
Proposed &
  \textbf{95.11\%} &
  \textbf{87.89\%} &
  \textbf{88.76\%} &
  \multirow{2}{*}{audiobook} \\ \cmidrule(r){1-4}
Transformer* &
  93.92\% &
  86.79\% &
  87.32\% &
   \\
BLSTM-CRF* &
  \multicolumn{1}{c}{94.41\%} &
  \multicolumn{1}{c}{86.74\%} &
  \multicolumn{1}{c}{87.15\%} &
   \\
Proposed* &
  \textbf{95.11\%} &
  \textbf{88.33\%} &
  \textbf{88.92\%} &
   \\ \bottomrule
\end{tabular}
\end{table}

\subsection{Objective evaluation results and analysis}
\label{sssec_r:results}

We choose F1 score as the evaluation metric. 
Experimental results are shown in Table \ref{table:experiment_main_table}.

On the MOOC dataset, Proposed outperforms BLSTM-CRF. 
Compared with BLSTM-CRF, F1 scores of Proposed on PW, PPH and IPH tasks are improved by 0.27\%, 0.91\% and 0.19\% respectively.
%On the audiobook dataset, Proposed also get better results than BLSTM-CRF. 
%Compared with BLSTM-CRF, Proposed has an absolute improvement of 0.70\%, 1.14\% and 1.05\% in terms of PW, PPH and IPH F1 scores. 
On the audiobook dataset, compared with BLSTM-CRF, Proposed has an absolute improvement of 0.70\%, 1.14\% and 1.05\% in terms of PW, PPH and IPH F1 scores. 
When the MTL framework is removed, Proposed* still achieves superior F1 scores than BLSTM-CRF* on the two datasets.
 %involving PW, PPH and IPH. 
%For ablation study, as the utterance encoder and the discourse encoder work together to extract inter-utterance linguistic information, we removed both components to further validate their effectiveness and the importance of multi-level contextual information for PSP.
%When both components are removed, Proposed becomes Transformer.
Compared with Proposed, F1 scores of PW, PPH and IPH prediction tasks of Transformer decrease 2.16\%, 2.18\% and 1.39\% on the MOOC dataset, and decrease 1.18\%, 1.02\% and 1.09\% on the audiobook dataset.
Moreover, when the MTL framework is removed, Transformer* still gets inferior performance than Proposed*.
The superior performances from both Proposed and Proposed* show that multi-level contextual information does help in all three subtasks of PSP, especially for PPH and IPH.
These experimental results also indicate the importance of the utterance encoder and the discourse encoder in our model design.

We also found that F1 scores of PPH and IPH from all models are consistently lower than PW in our experiments, which is also a common phenomenon in previous works \cite{pan2019mandarin,zheng2018blstm,lu2019self,zhang2020unified}.
We inspected precision and recall of all models on test set for PPH and IPH, and found that recall is lower than precision.
According to Table \ref{table:statistics_datasets}, this can be explained by the imbalanced dataset where positive labels are significantly less than negative ones for PPH and IPH.
We will explore solutions for this issue in future work. 

\subsection{Experiments on context window size}

We also conduct experiments to determine the best context window size.
Results are shown in Table \ref{table:window_size}.
Generally speaking, a larger window brings better prediction performance, but with diminishing returns.
Note that when the size of the context window is set to 1, Proposed becomes a PSP model which has more parameters but doesn't employ inter-utterance linguistic information, and it has the worst prediction performance.
On two datasets, the prediction performance begins to saturate when context window size is larger than 8.
%On MOOC dataset the prediction performance of our model stops growing when we increase the context window size from 8 to 16 , while on audiobook dataset using a context window of size 16 still improves overall prediction performance.
%Thus, different datasets should use different sizes of context window.

We should also notice that bigger context window will introduce extra computation overhead during training.
Assume $n$ is the size of the context window, $N$ is the number of utterances in the training corpus.
Each training sample in the training set is $C_p$ containing $n$ utterances, where $p \in [1,2,3,...,N-n+1]$.
Thus, computation for each training sample grows linearly with the increase of context window size.
To better balance prediction performance between computation overhead of training, the size of the context window for Proposed is set to 8 in our experiments.

\begin{table}
\caption{Experiments on context window size for Proposed on MOOC and audiobook datasets.}
\label{table:window_size}
\begin{tabular}{@{}clllc@{}}
\toprule
\multicolumn{1}{l}{\multirow{2}{*}{\textbf{Window Size}}} &
  \multicolumn{3}{c}{\textbf{F1 score}} &
  \multirow{2}{*}{\textbf{Dataset}} \\ \cmidrule(lr){2-4}
\multicolumn{1}{l}{} &
  \multicolumn{1}{c}{\textbf{PW}} &
  \multicolumn{1}{c}{\textbf{PPH}} &
  \multicolumn{1}{c}{\textbf{IPH}} &
   \\ \midrule
1  & 94.47\%          & 73.20\%          & 76.57\%          &                       \\
2  & 94.83\%          & 73.99\%          & 76.99\%          &                       \\
4  & 94.95\%          & 74.63\%          & 77.18\%          & \multirow{2}{*}{MOOC} \\
8  & 95.35\%          & \textbf{75.15\%} & 77.89\%          &                       \\
12 &
  \multicolumn{1}{c}{95.31\%} &
  \multicolumn{1}{c}{75.01\%} &
  \multicolumn{1}{c}{77.77\%} &
  \multicolumn{1}{l}{} \\
16 & \textbf{95.39\%} & 74.94\%          & \textbf{77.95\%} &                       \\ \midrule
1  & 94.45\%          & 87.30\%          & 88.19\%          &                       \\
2  & 94.83\%          & 87.63\%          & 88.79\%          &                       \\
4 &
  94.99\% &
  87.82\% &
  \textbf{88.94\%} &
  \multirow{2}{*}{audiobook} \\
8  & 95.11\%          & \textbf{87.89\%} & 88.76\%          &                       \\
12 &
  \multicolumn{1}{c}{\textbf{95.20\%}} &
  \multicolumn{1}{c}{87.81\%} &
  \multicolumn{1}{c}{88.66\%} &
  \multicolumn{1}{l}{} \\
16 & 95.17\%          & 87.85\%          & 88.76\%          &                       \\ \bottomrule
\end{tabular}
\end{table}

\subsection{Subjective evaluation results and analysis}

\begin{figure}[!htb]
\centering
\includegraphics[width=1\linewidth]{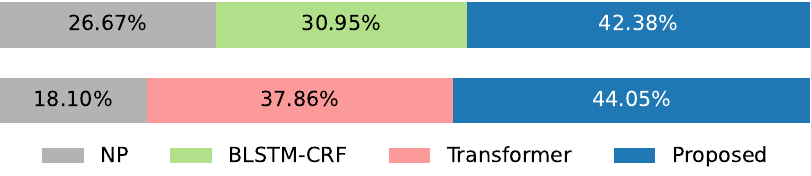}
\caption{Results of ABX tests for naturalness of synthesized speech on audiobook dataset. NP means no preference.}
\label{fig:abx}
\end{figure}

To further evaluate the PSP performance of different methods and their impact on TTS synthesis, we further conduct ABX preference tests on the naturalness of the synthesized speech.
Because audios from the MOOC dataset contain too much noise to train a TTS model, we only conduct ABX preference tests on the audiobook dataset.
30 samples that are longer than 30 characters are randomly selected from the test set with different prosodic structures predicted by aforementioned different models and the corresponding speeches are generated using Huya TTS toolkit, which is based on DurIAN \cite{durian} and  HiFi-GAN \cite{hifigan}.
%zywu-0327
%重写这部分
%subject怎么打分？ABX，NP啥意思
%多少subjects参与打分
The synthesized speeches corresponding to different models are presented to subjects in random order.
14 subjects were asked to choose a preferred speech in terms of naturalness of the speeches for each speech pair.
%Each utterance  was presented to volunteers, who are asked to choose a preferred speech in terms of naturalness of the speech.
Results are shown in Fig \ref{fig:abx},
which demonstrate that compared with BLSTM-CRF and Transformer, our proposed method leads to higher naturalness of the synthesized speech.

\section{Conclusions}

In this paper, we propose a model utilizing multi-level contextual information, which includes both intra-utterance and inter-utterance linguistic information, to improve the performance of PSP. 
This model consists of a hierarchical encoder and a MTL decoder. 
Objective experimental results on two datasets show that compared with baseline models, our proposed method achieves better F1 scores on PW, PPH and IPH prediction tasks.
%Objective experimental results on two datasets show that our proposed method achieves better F1 scores on PW, PPH and IPH prediction tasks.
We also performed ABX preference tests to evaluate the improvement brought by our method on synthesized speeches, where our method again outperforms baseline models.
%Objective experimental results and subjective experimental results demonstrates the effectiveness of our proposed method.

\textbf{Acknowledgement}: This work was supported by National Key R\&D Program of China (2020AAA0104500), National Natural Science Foundation of China (NSFC) (62076144) and National Social Science Foundation of China (NSSF) (13\&ZD189).

\bibliographystyle{IEEEtran}

\bibliography{mybib}

\end{document}